\documentclass[aps,prc,reprint,groupedaddress,amsmath]{revtex4-1}

\usepackage{graphicx}
\usepackage{braket}
\usepackage{bm}
\usepackage{longtable}

\begin{document}

\title{Cluster states and isoscalar monopole transitions of $^{24}$Mg}

\author{Y. Chiba}
\author{M. Kimura}
\affiliation{Department of Physics, Hokkaido University, Sapporo 060-0810, Japan}

\date{\today}

\begin{abstract}
\begin{description}
\item[Background] Isoscalar monopole transition has been suggested as a key
  observable to search for exotic cluster states. Recently, the excited
  $0^+$ states with strong isoscalar monopole transition strengths are
  experimentally reported in $^{24}$Mg, but their structures are unrevealed
  because of the lack of theoretical analysis.
\item[Purpose] Study structure of the excited $0^+$ states of $^{24}$Mg
  populated by isoscalar monopole transition from the ground state and identify
  their cluster configurations.
\item[Method] The $0^+$ states of $^{24}$Mg and their isoscalar monopole
  transition strengths from the ground state are calculated with
  antisymmetrized molecular dynamics combined with generator coordinate method
  using Gogny D1S interaction.
\item[Results] The calculated isoscalar monopole strength function shows
  reasonable agreement with experiment and is consistent with other
  theoretical calculation. The structure of the excited $0^+$ states with
  pronounced isoscalar monopole transitions are analyzed. It is found that the
  $0^+_2$, $0^+_3$ and $0^+_5$ states have mixed nature of mean-field and
  cluster, and that the $0^+_8$ state is dominated by $^{12}$C+$^{12}$C
  cluster configuration. In addition, it is predicted that $5\alpha$-pentagon+$\alpha$ states appear around 23 MeV.
\item[Conclusions] The excited $0^+$ states which appear as the prominent peaks
  in the calculated strength function are associated with $^{20}$Ne+$\alpha$,
  $^{12}$C+$^{12}$C and $5\alpha$-pentagon+$\alpha$ cluster states.
\end{description}
\end{abstract}

\pacs{Valid PACS appear here}
\maketitle

{\it Introduction.---} Clustering is a fundamental degree-of-freedom of nuclear
excitation. According to Ikeda threshold rule \cite{IkedaPTPE68}, the appearance of
various cluster states is expected near the cluster decay thresholds.
Clustering of $p$ shell nuclei has long been studied and well
established \cite{Fujiwara,Freer} including the dilute gas-like $\alpha$-cluster state of
$^{12}$C($0^+_2$) \cite{UegakiPTP57,
KamimuraNPA351,EnyoPRL81,THSR,FunakiPRC,NeffPRL98}. On the other hand, in the mid
$sd$-shell nuclei, the existence of cluster states is not well established,
although many interesting phenomena can be expected. For example, in the case
of $^{24}$Mg, a variety of exotic cluster states is expected: In addition to
the ordinary $\alpha$ cluster state ($^{20}$Ne+$\alpha$), $^{12}$C+$^{12}$C
molecular states of astrophysical interest
\cite{BromleyPRL,AlmqvistPRL,ImanishiNPA125,Kondo,OhkuboPTP67,DesBayeCC,KatoPTP81,Betts,HoNature},
$^{16}$O$+2\alpha$ clustering \cite{DesBaye,KatoNPA463,ItagakiPRC75,IchikawaPRC83} and 6$\alpha$
condensation \cite{THSR,YamadaPRC69,Girod01} are theoretically discussed. However, their high
excitation energies make it difficult to identify them experimentally.

In this decade, it is found that the isoscalar (IS) monopole transition strengths
between the ground and excited cluster states are considerably enhanced,
and hence, it can be a good probe for highly excited cluster states. The
discussion was initiated by T. Kawabata \cite{KawabataB} and Y. Kanada-En'yo \cite{EnyoB} on the enhanced IS
monopole transition of $^{11}$B between the shell model like ground state and
the $3/2^-_3$ state with pronounced $2\alpha+t$ cluster structure.
 T.~Yamada {\it et al.} \cite{YamadaPTP120,YamadaPRC85} proved the mechanism
of the enhanced IS monopole transition using cluster-model wave function. The
ingredient of the enhancement is the fact the ground state has ``{\it duality nature}'' of the mean-field and
clustering \cite{PerringSkyrme,BaymanBohr}. The duality nature implies
that the degrees-of-freedom of cluster excitation is embedded in the ground
state even if it has a pure shell-model structure. It was shown that the IS
monopole transition operator can activate this degrees-of-freedom of
clustering. As a result, the excited cluster states can be strongly populated
by the IS monopole transitions. In fact, the enhancements of IS monopole
transition strengths are observed in $p$-shell nuclei such as $^{11}$B
\cite{KawabataB}, $^{12}$C \cite{AjzenbergNPA506} and $^{16}$O
\cite{AjzenbergNPA460}, and they nicely coincide with the cluster states
predicted by theoretical calculations. Thus, IS monopole transition is a
promising probe for highly excited cluster states. 

Recently, the excited $0^+$ states with strong isoscalar monopole transition
strengths are experimentally reported in $^{24}$Mg \cite{KawabataMg}, but their
structures are ambiguous. Therefore, in this study, we aim to clarify the
relationship between those excited $0^+$ states and clustering. For this
purpose, we calculate the excited $0^+$ states of $^{24}$Mg and investigate
their clustering and IS monopole transition strengths from the ground state by
the antisymmetrized molecular dynamics (AMD), which has successfully
described a variety of structure of $p$-$sd$-$pf$-shell nuclei \cite{AMDCRP,AMDPTP,ChibaPRC}
including of the low-lying states of $^{24}$Mg \cite{KimuraPTP127}. To
describe the cluster states and single-particle states including giant monopole
resonance (GMR) simultaneously, we introduce the constraint on the harmonic
oscillator quanta and perform the generator coordinate method (GCM) with a
large number of basis wave functions.

{\it Formalism.---}We employ the microscopic Hamiltonian,
\begin{align}
 \hat{H} &= \sum_i^A{\frac{\hat{p}_i^2}{2m}}
 - \hat{t}_{c.m.} + \sum_{i < j}^A{\hat{v}_{NN}(ij)} + \sum_{i < j}^Z{\hat{v}_{C}(ij)}, 
\end{align}
where $\hat{t}_{c.m.}$, $\hat{v}_{N}$ and $\hat{v}_{C}$ stand for the center-of-mass kinetic
energy, Gogny D1S effective NN interaction \cite{BergerCPC63} and the Coulomb interaction approximated
by a sum of seven Gaussians, respectively. The AMD variational wave function used in this study is an antisymmetrized
product of the single particle wave packets projected to the positive-parity state,
\begin{align}
 \Phi^+ &= \frac{1+\hat{P}_x}{2} \mathcal{A}\left\{ \varphi_1,\varphi_2,\dots \varphi_A \right\},\\
 \varphi_i(\bm{r}) &= \exp{\left[- \sum_{\sigma = x,y,z}{\nu_{\sigma}
 \left( r_\sigma - \frac{Z_{i\sigma}}{\sqrt{\nu_\sigma}} \right)^2}
 \right]} \\
 & \otimes (a_i \chi_{\uparrow} + b_i \chi_{\downarrow}) \otimes (\text{neutron or proton}),
 \label{DEF:HamiltonianAndVwf}
\end{align}
where the single-particle wave packet $\varphi_i$ is represented by a deformed Gaussian
wave packet \cite{KimuraPRC69}, and the variational parameters $\nu_\sigma$, $\bm{Z}_{i}$, $a_i$
and $b_i$ are determined by the energy variation. 

To deal with the low-lying quadrupole collective states and highly excited
cluster states simultaneously, we introduce two different constraints in the
energy variation. The first is imposed on the nuclear quadrupole deformation
parameters $\beta$ and $\gamma$ to describe the low-lying collective states,
and we denote the set of the wave functions obtained with this constraint as
$\Phi^+_{\beta\gamma}$. As the second constraint, we extend the method used in
Ref. \cite{EnyoPRC72} and impose the constraint on the expectation values of the
harmonic oscillator quanta $N_x$, $N_y$ and $N_z$, which are defined as the
eigenvalues of the 3 by 3 matrix, 
\begin{align}
 N_{\sigma\tau} &= \langle\Phi^+|\sum_{i=1}^A a_\sigma^\dagger(i) a_\tau(i)|\Phi^+ \rangle,
 \quad \sigma, \tau = x,y,z.
\end{align}
Here $a_\tau(i)$ is an ordinary annihilation operator of the harmonic
oscillator acting on the $i$th nucleon, and the oscillator parameter
$\hbar\omega$ is estimated from the ground state radius and set to 12.6 MeV. As
a measure of the particle-hole excitation, we introduce the quantity $\Delta N
= N_x+N_y+N_z-N_0$ where $N_0$ is the lowest Pauli-allowed value equal to 28.
Under the condition of the $\Delta N=0, 2, 4, 6$ or 8, we put the constraints
for all possible even integer values of $N_x, N_y$ and $N_z$. In other words,
roughly speaking, we searched for the various many-particle-hole
configurations within 8$\hbar\omega$ excitation. We denote thus-obtained set of
the wave functions as $\Phi^+_{\Delta N}$. 

We further introduce an additional set of the basis wave functions $\Phi^+_{IS0}$ defined as, 
\begin{align}
  \Phi^+_{IS0} &= \left( 1 - e^{-\mu\hat{O}_{IS0}} \right) \Phi^+_{\beta\gamma} 
  \simeq \mu \hat{O}_{IS0} \Phi^+_{\beta\gamma}, \\
 \hat{O}_{IS0} &= \sum_{i=1}^{A} {(\bm{r}_i-\bm{r}_{c.m.})^2},
  \label{Def:monopole-basis}
\end{align}
where $\mu$ is arbitrary small real number, $\hat{O}_{IS0}$ is the IS monopole operator
and $\bm r_{c.m.}$ is the center-of-mass coordinate. By definition, the set of the wave functions
$\Phi^+_{IS0}$ describes $1p1h$ $(2\hbar\omega)$ excited states built on $\Phi^+_{\beta\gamma}$ by 
the IS monopole operator. The similar method was also used in Ref.~\cite{WHoriuchiPRC85}.

Those three sets of wave functions $\Phi^+_{\beta\gamma}$, $\Phi^+_{\Delta N}$ and $\Phi^+_{IS0}$
are projected to the $J^\pi=0^+$ and superposed to describe various $0^+$ states from the
low-lying to the highly excited states (GCM),
\begin{align}
 \Psi^{0^+}_{n} = \sum_{ i\in \Phi^+_{\beta\gamma}, \Phi^+_{\Delta N}, \Phi^+_{IS0}} 
 {g_{in} P^{J=0}\Phi^+_i}, 
  \label{Def:GCMWf}
\end{align}
where $P^{J=0}$ is the projector to $J=0$ state. We superposed 524 basis wave function $\Phi_i^+$
in total, and solved the Hill-Wheeler equation to obtain the eigenenergies $E_{n}$ and wave
functions $\Psi_n^{0^+}$ of the ground and excited $0^+_n$ states. To discuss the intrinsic
structure of the $0^+$ states, we also calculate the overlaps between each $0^+$ state and the
basis wave functions, $|\langle\Psi^{0^+}_n|P^{J=0}\Phi^{+}_i\rangle|^2 / \langle P^{J=0}\Phi^+_i|P^{J=0}\Phi^+_i\rangle $.

Using the wave functions of the ground and excited $0^+$ states directly, we derived the IS
monopole matrix elements $M_n(IS0)$, reduced transition strengths $B(IS0)$, strength function
$S(E_x)$ and the energy non-weighted and weighted sums $m_k$ with $k=0,1,3$, 
\begin{align}
  &M_n(IS0) = \braket{\Psi^{0+}_n|\hat{O}_{IS0}|\Psi^{0+}_{g.s.}},\\
  &B(IS0;g.s.\rightarrow 0^+_n) = |M_n(IS0)|^2, \\
  &S(E_x) = \sum_n |M_n(IS0)|^2 E'_n \delta(E'_n-E_x), \\
  &m_k = \int^{\infty}_{0} dE_x \sum_n |M_n(IS0)|^2 E'^k_n \delta(E'_n-E_x),
  \label{Def:StrengthFunction}
\end{align}
where $E'_n$ stands for the excitation energy of the $n$th $0^+$ state, 
{\it i.e.} $E'_n=E_n - E_{g.s.}$.
\begin{figure}[b]
  \includegraphics[width=\hsize]{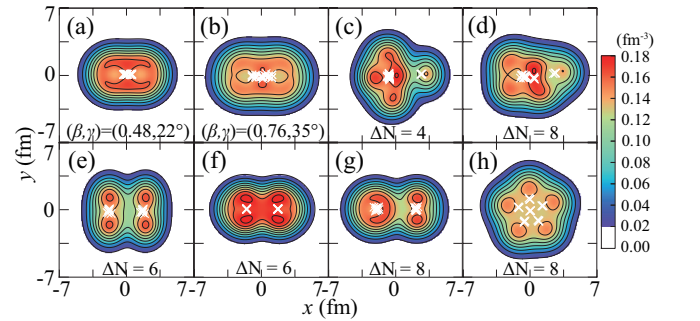}
  \caption{Intrinsic density distributions at the $z=0$ plane obtained by constraint on the
 matter quadrupole deformation parameters ((a) and (b)) and the expectation values of the
 harmonic oscillator quanta ((c)-(h)). The crosses in each figure show the centroids 
 of Gaussians describing nucleons. The contour lines are plotted in the interval of 0.02 fm$^{-3}$}\label{fig:den}
\end{figure}

{\it Results of the energy variation.---} 
Figure \ref{fig:den} (a) and (b) show the typical configurations obtained by the constraint on the
quadrupole deformation. After the GCM calculation, they become the dominant component of the
ground and $0^+_2$ states, respectively. The centroids of the Gaussian wave packets are gathered
around the center-of-mass, describing triaxially deformed mean-field configuration. As
already discussed in our previous work \cite{KimuraPTP127}, the constraint on the quadrupole deformation
generates deformed mean-field configurations \cite{RodriguezPRC81,BenderEPJA17}, but no cluster configuration. 

The use of the constraint on the harmonic oscillator quanta generates various kind of cluster
configurations as well as single-particle excited configurations with approximate
$\Delta N\hbar\omega$ excitations which lie energetically above the energy surface of
$\Phi_{\beta\gamma}$ and are not accessible by the constraint on the quadrupole deformation. The
panels (c)-(h) show the examples of thus-obtained cluster wave functions, and they are the
dominant component of the excited $0^+$ states corresponding to the prominent peaks in the IS
monopole strength function $S(E_x)$. By the constraint of $\Delta N=2$, $^{20}{\rm Ne}+\alpha$ and
$^{12}{\rm C}+{}^{12}{\rm C}$ cluster states start to appear. As $\Delta N$ increases, the
inter-cluster distance grows and the orientation of cluster changes depending on the combination
of $N_x, N_y$ and $N_z$. For example, the panels (c) and (d) show the $^{20}{\rm Ne}+\alpha$
cluster configuration with $\Delta N=4$ and 8, which mainly contribute to the $0^+_2$ and $0^+_5$
states, respectively. They have different orientation of $^{20}$Ne cluster and inter-cluster
distances (distance between the centroids of Gaussians describing clusters) are 3.0 and 3.3 fm,
respectively. The panels (e), (f) and (g) show $^{12}$C+$^{12}$C cluster states with $\Delta N=6$,
6 and 8, respectively. They have different orientations of the oblately deformed $^{12}$C
clusters, and inter-cluster distances are 3.5, 3.5 and 4.0 fm. By further increase of $\Delta N$,
very exotic cluster structure composed of 6$\alpha$ particles appears. A typical example is shown in the panel (h) which was obtained by the constraint
of $\Delta N=8$. In this configuration, centroids of Gaussians describing
5$\alpha$ clusters locate at the vertex of a pentagon with side of 1.5 fm, and
the last $\alpha$ cluster is 0.25 fm above it. After the GCM calculation, this
kind of 5$\alpha$-pentagon + $\alpha$ configurations generate two $0^+$ states
above 20 MeV. Thus, by increasing the number of particle-hole, $^{24}$Mg is
clustered as illustrated in Ikeda diagram \cite{IkedaPTPE68}.

\begin{figure*}[t]
  \includegraphics[width=\hsize]{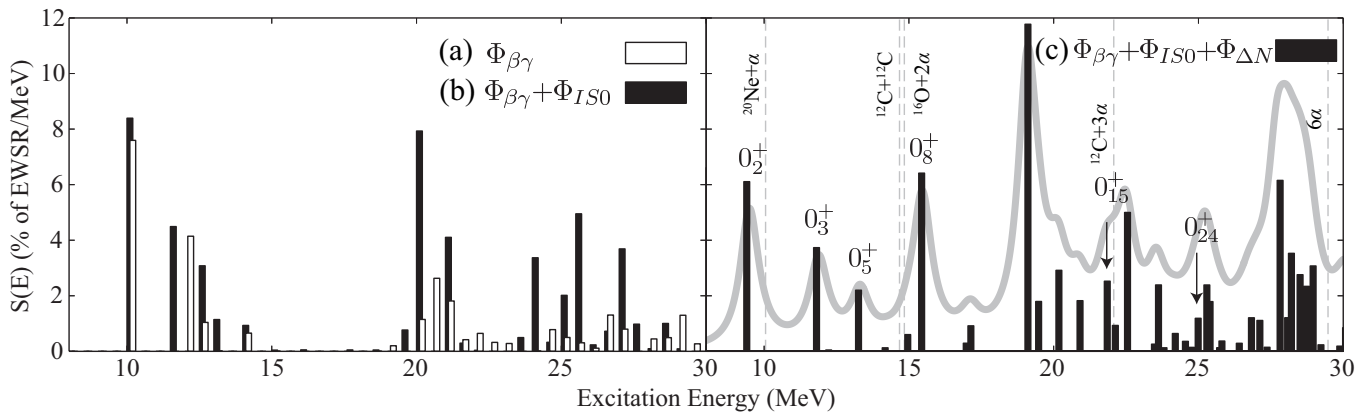}
  \caption{The isoscalar monopole transition strength functions calculated with the 
    basis sets of (a)
  $\Phi_{\beta\gamma}$, (b) $\Phi_{\beta\gamma}+\Phi_{IS0}$ and (c)
  $\Phi_{\beta\gamma}+\Phi_{IS0}+\Phi_{\Delta N}$. The solid line in the
  right panel shows the strength function smeared by Lorentzian 
  with 0.8 MeV width. The vertical dashed lines indicate cluster decay threshold
  energies which are located at the observed binding energies.}\label{fig:ismsf}
\end{figure*}

{\it IS monopole transition strengths.---} 
\begin{table}
 \centering
 \caption{Calculated energy weighted sums $m_1$ and $m_1^*$ in fraction of the
   EWSR and the centroid energies of GMR ($m_1^*/m_0^*$
   and $\sqrt{m_3^*/m_1^*}$) in MeV, where $m^*_0$, $m_1^*$ and $m_3^*$ are the sums
   between $E_x=$ 9 and 40 MeV excluding the $0^+_2$ state.
    }\label{tab:sum}
  \begin{ruledtabular}

  \begin{tabular}{lcccc} 
   basis set  & $m_1$ &$m_1^*$   & $m_1^*/m_0^*$       & $\sqrt{m_3^*/m_1^*}$ \\ \hline
   (a) $\Phi_{\beta\gamma}$  & 35  &26 & 20.3 & 24.2   \\
   (b) $\Phi_{\beta\gamma}$+$\Phi_{IS0}$  & 116 &101  & 25.6  & 29.3  \\
   (c) $\Phi_{\beta\gamma}$+$\Phi_{IS0}$+$\Phi_{\Delta N}$ & 103 &90  & 22.2 & 25.2  \\
   \hline
   exp. \cite{YoungbloodPRC60,XChenPRC80,YoungbloodPRC80}   &    & 82 $\pm$ 9 & 21.9$^{+0.3}_{-0.2}$ & 24.7$^{+0.5}_{-0.3}$      \\
   QRPA \cite{PeruPRC77}   &    & 94         & 20.57                &                           \\
  \end{tabular}
  \end{ruledtabular}
\end{table}
The ground and excited $0^+$ states are calculated by the GCM with three
different basis sets (a) $\Phi_{\beta\gamma}$ (b)
$\Phi_{\beta\gamma}+\Phi_{IS0}$ (c) $\Phi_{\beta\gamma}+\Phi_{IS0}+\Phi_{\Delta
N}$. The IS monopole transition strengths derived from these GCM wave functions
are shown in Fig. \ref{fig:ismsf}, and their energy weighted sums and the
centroid energies of GMR are summarized in Tab. \ref{tab:sum}. With only the
basis set $\Phi_{\beta\gamma}$, the strength function (Fig. \ref{fig:ismsf}
(a)) fails to describe GMR, and the energy weighted sum $m_1$ amounts to only
35\% of the sum rule (EWSR). Addition of the basis set $\Phi_{IS0}$ (Fig.
\ref{fig:ismsf} (b)) greatly improves $m_1$ value (116\% of EWSR), but
overestimates the observed GMR centroid energy \cite{YoungbloodPRC60,
XChenPRC80,YoungbloodPRC80} because the GMR strength
distributes widely in the region of $E_x > 30$ MeV. The inclusion of the basis
set $\Phi_{\Delta N}$ yields the reasonable strength function as shown in Fig.
\ref{fig:ismsf} (c). Namely various cluster and single-particle states with
$\Delta N\hbar \omega$ excitation described by $\Phi_{\Delta N}$ lower the GMR
position and enhance its strength. As a result, the strength function exhausts
approximately 100\% of EWSR and plausibly agrees with the experimental energy
weighted sum and the GMR centroid energy observed in the energy range of $E_x =
9 - 40$ MeV. It is also noted that the quasi-particle random phase
approximation (QRPA) with Gogny D1S interaction \cite{BergerCPC63} also yielded similar values and
qualitatively agrees with our results and experiment with respect to the
global structure of GMR.

From the comparison between the strength functions shown in Fig.
\ref{fig:ismsf} (b) and (c), we also see that not only the GMR strength
($E_x\gtrsim 18$ MeV) but also the low-lying structure ($E_x\lesssim 18$ MeV)
of the strength function is largely modified by the basis set $\Phi_{\Delta
N}$. For example, note that the prominent peak at 15.3 MeV in Fig.
\ref{fig:ismsf} (c) is completely missing in Fig. \ref{fig:ismsf} (b). Based on
the analysis of the wave functions corresponding to those peaks, we conclude
that several prominent peaks are attributed to the cluster configurations and
suggest that the cluster states shown in Fig. \ref{fig:den} can be populated
and observed by their enhanced IS monopole transition strengths. To see this
point, we discuss the structure of the $0^+$ states relevant to the prominent
peaks in $S(E_x)$ in the following.

{\it Cluster states and their transition strengths.---} 
The ground state is dominated by the mean-field structure and has the
largest overlap (0.93) with the wave function shown in Fig.~\ref{fig:den}~(a).
However, at the same time, it also has non-negligible overlaps with the cluster
wave functions. It has 0.26 and 0.40 overlaps with
$^{20}$Ne+$\alpha$ and $^{12}$C+$^{12}$C cluster states shown in
Fig.~\ref{fig:den}~(c) and (e), respectively. This result means following two
points. The first is that the cluster correlation exists even in the ground
state. The binding energy of the ground state increases from 198.3 MeV to 199.2
MeV by including $\Phi_{\Delta N}$ which indicates that the additional binding
energy of 0.9 MeV is brought about by the cluster correlation. Secondly, it
shows that the ground state has ``{\it duality nature}'' of the mean-field and
clusters and that the degrees-of-freedom of cluster excitation are embedded in
the ground state. This is an essential ingredient for the discussion of the
IS monopole transition from the ground to the excited cluster states
\cite{YamadaPTP120}.

By including $\Phi_{\Delta N}$, the low-lying mean-field states (the excited
states having $E_x < 15$ MeV in Fig.~\ref{fig:ismsf}~(b)) are strongly mixed
with $^{20}{\rm Ne}+\alpha$ and $^{12}$C+$^{12}$C cluster states and constitute
the low-lying prominent peaks at 9.3, 11.7 and 13.2 MeV in
Fig.~\ref{fig:ismsf}~(c), which correspond to the $0^+_2$, $0^+_3$ and $0^+_5$
states, respectively. In contrast to those mixed states, the $0^+_8$ state at
15.3 MeV is dominated by the $^{12}$C+$^{12}$C cluster configurations.
Furthermore, 5$\alpha$-pentagon + $\alpha$ cluster states configurations
generates the $0^+_{15}$ and $0^+_{24}$ states at 21.8 and 24.9 MeV. 

The $0^+_2$ state which appears as the lowest peak at 9.3 MeV has the largest
overlap (0.36) with the mean-field configuration of Fig.~\ref{fig:den}~(b)
which has larger quadrupole deformation parameter $\beta$ than the ground
state. It can be regarded as the $\beta$-band built on the ground band, and
hence, has large IS monopole transition strength as listed in
Tab.~\ref{tab:prop}. However, it also has 0.32 overlap with the $^{20}{\rm
Ne}+\alpha$ cluster configuration shown in Fig.~\ref{fig:den}~(c). Owing to
this cluster correlation, it gains additional binding energy of 1.8 MeV which
reduces the excitation energy from 10.2 MeV to 9.3 MeV.
The $^{20}$Ne+$\alpha$ cluster structure also constitute the $0^+_5$ state at 11.7 MeV by 
the mixing with mean-field structure. It has the largest overlap (0.30)
with the configuration of the $^{20}$Ne+$\alpha$ cluster configuration shown in
Fig.~\ref{fig:den}~(d) and the comparable overlap (0.25) with the mean-filed
wave function with $(\beta,\gamma)=(0.4, 57^\circ)$ in $\Phi_{\beta\gamma}$.

\begin{table}
  \centering
  \caption{Properties of the $0^+$ states obtained by GCM calculation with
  $\Phi_{\beta\gamma}$, $\Phi_{IS0}$ and $\Phi_{\Delta N}$. $E_x$,
proton radius $\sqrt{\braket{r^2_p}}$, $B(IS0)$ and $E_x B(IS0)$ are given in
unit of MeV, fm, fm$^4$ and fraction of EWSR in percentage, respectively. The
values in bracket are the observed values
\cite{StrehlZPhys214,EndtANDT,Yordanov,CODATA}. The observed $\sqrt{r^2_p}$ is
deduced from the observed charge radius \cite{Yordanov} and the proton charge
radius \cite{CODATA}.}\label{tab:prop}
  \begin{ruledtabular}
  \begin{tabular}{ccccc} 
    State                        & $E_x$       & $\sqrt{\braket{r^2_p}}$ & $B(IS0)$         & $E_x B(IS0)$ \\ \hline
    $0^+_1$                      & 0.0         & 3.06 (2.93)             &                  &              \\
    $0^+_2$                      & 9.3 (6.4)   & 3.11                    &  122 ($180 \pm 20$)    &  6.1 ($6.4 \pm 0.7$) \\
    $0^+_3$                      & 11.7        & 3.08                    &  59.3            &  3.7         \\
    $0^+_5$                      & 13.2        & 3.06                    &  31.1            &  2.2         \\ 
    $0^+_8$                      & 15.3        & 3.11                    &  77.8            &  6.4         \\
    $0^+_{15}$                   & 21.8        & 3.14                    &  21.6            &  2.5         \\
    $0^+_{24}$                   & 24.9        & 3.28                    &  8.90            &  1.2         \\
  \end{tabular}
  \end{ruledtabular}
\end{table}

The $^{12}$C+$^{12}$C cluster configurations dominantly contribute to the
$0^+_3$ and $0^+_5$ states. The $0^+_3$ state at 11.7 MeV exhausts 3.7 \% of
EWSR and has the large overlaps with $^{12}$C+$^{12}$C, $^{20}$Ne+$\alpha$
cluster and mean-field configurations. The overlaps are 0.21, 0.19 and 0.16
with $^{12}$C+$^{12}$C and $^{20}$Ne+$\alpha$ cluster and the mean-field
configurations with $(\beta,\gamma) = (0.76,2.4^\circ)$, respectively. In
contrast to the above mentioned states, the $0^+_8$ state at $15.3$ MeV, which
is close to the $^{12}$C+$^{12}$C cluster decay threshold energy, is governed
by the $^{12}$C+$^{12}$C cluster configurations. It has 0.14 0.11 0.18 overlaps
with the configurations of Fig.~\ref{fig:den}~(e), (f) and (g). The overlaps
with other configurations are less than 0.09. The $0^+_8$ state has
strong IS monopole transition strength and 6.4 \% of EWSR. It is noted that
this state is completely missing in Fig.~\ref{fig:ismsf}~(a), (b) and looks also
missing in the QRPA calculation \cite{PeruPRC77}, which is consistent with its
strong $^{12}$C+$^{12}$C cluster nature. Very interestingly, the $0^+$ state
with enhanced IS monopole transition is observed at the $^{12}$C+$^{12}$C
cluster threshold energy in the $^{24}$Mg$(\alpha,\alpha')$ experiment
\cite{KawabataMg}, which can be associated with the present $0^+_8$ state. We
also note that the $0^+_8$ state is the band-head of $^{12}$C+$^{12}$C cluster
band which is a candidate of the observed $^{12}$C+$^{12}$C molecular
resonances of astrophysical interest, for which we will discuss in detail in our
forthcoming paper.

Adding to those clusters, the 5$\alpha$-pentagon+$\alpha$ cluster states appear as the
$0^+_{15}$ and $0^+_{24}$ states at 21.8 and 24.9 MeV that exhaust 2.5 and 1.2
\% of EWSR, respectively. The $0^+_{15}$ state has the largest overlap which
amounts to 0.43 with the 5$\alpha$-pentagon+$\alpha$ cluster structure shown
in Fig.~\ref{fig:den}~(h). It also has non-negligible overlap (0.20) with the
single-particle excited configuration described by $\Phi_{IS0}$. The $0^+_{24}$
state has the largest overlap (0.26) with the similar configuration to that of
Fig.~\ref{fig:den}~(h) and has rather minor contributions from other
configurations. One may attempt to associate these 5$\alpha$-pentagon+$\alpha$
cluster states with dilute $6\alpha$ gas state analogous to the Hoyle state of
$^{12}$C($0^+_2)$. Indeed, the recent Hatree-Fock-Bogoliubov calculation showed
the possible existence of dilute $n\alpha$ cluster states at very low density
in $N=Z$ nuclei \cite{Girod01}. However, we conclude that the calculated $0^+_{15}$
and $0^+_{24}$ states do not correspond to dilute $\alpha$ gas state suggested
by the HFB calculation according to the following reasons. First, the radii of
those states are not large enough to be a dilute $\alpha$ gas state and much
smaller than those of Ref.~\cite{Girod01}. Second, they are much more deeply
bound compared to $E_x = 80$ MeV reported in Ref.~\cite{Girod01}. Since AMD is
free from the spurious center-of-mass kinetic energy and the parity and
angular-momentum projections are correctly performed, the excitation energies
of cluster states are greatly reduced. In addition, the present
$0^+_{15}$ and $0^+_{24}$ states are rather compact and have non-negligible
interaction energies between $\alpha$ clusters. Therefore, we conclude that
those states have cluster structure but do not have dilute gas nature. We
conjecture that dilute 6$\alpha$ gas state will appear at higher excitation
energy and to describe it, we will need to enlarge $\Delta N$ in
order to include more spatially extended 6$\alpha$ configurations.
Nevertheless, we emphasize that the exotic $\alpha$ cluster states are firstly
obtained without {\it a priori} assumption on clustering in this study and it is shown that
they are experimentally accessible via IS monopole transition from the ground
state. Although they are embedded in the GMR energy region, we expect that
they can be experimentally identified by their decay mode, because different
from GMR, they will selectively decay through $\alpha$ particle emission.

{\it Conclusions.---} In summary, we investigated the structure of the excited
$0^+$ states of $^{24}$Mg and their IS monopole transition strengths based on
AMD. The mean-field and cluster configurations of $^{24}$Mg were
obtained by the energy variation. In particular, by using the constraint on
the harmonic
oscillator quanta, the $^{20}$Ne+$\alpha$,
$^{12}$C+$^{12}$C, and 5$\alpha$-pentagon+$\alpha$ cluster configurations were
obtained without any {\it a priori} assumption on clustering. In addition,
$1p1h$ $(2\hbar\omega)$ excited configurations built 
 by the IS monopole operator were also introduced
 as the basis wave functions of GCM. With these
 basis wave functions, the calculated $0^+$ states yielded reasonable IS monopole strength function.
Namely, they exhausted almost 100 \% of EWSR
and reproduced the observed centroid energy of GMR. The result is
also consistent with the QRPA calculation.

We have shown that the several excited $0^+$ states with the enhanced IS monopole transitions
are associated with $^{20}$Ne+$\alpha$,
$^{12}$C+$^{12}$C and $5\alpha$-pentagon+$\alpha$ cluster configurations.
The $0^+_2$, $0^+_3$ and $0^+_5$ states have the mixed
nature of mean-field, $^{20}$Ne+$\alpha$ and $^{12}$C+$^{12}$C cluster
configurations, while the $0^+_8$ state is governed by
$^{12}$C+$^{12}$C cluster configuration. 
The $0^+_8$ state may be associated with the strong peak observed at 
the $^{12}$C+$^{12}$C cluster threshold energy in the
$^{24}$Mg$(\alpha,\alpha')$ experiment \cite{KawabataMg}.
Furthermore, we predicted that the 5$\alpha$-pentagon+$\alpha$ cluster states
appear in the GMR energy
region. Even though they do not correspond to the dilute 6$\alpha$ gas state,
it is emphasized that the exotic $\alpha$ cluster states were firstly obtained
without any {\it a priori} assumption on clustering and were shown to to be
experimentally accessible via IS monopole transition from the ground state. We
expect that the detailed comparison with the latest experimental data will
reveal the exotic clustering of $^{24}$Mg.

The authors acknowledge Prof. T. Kawabata, Prof. Y. Taniguchi and Y. Funaki for
the valuable discussions. Part of the numerical calculations were performed on
the HITACHI SR16000 at KEK. One of the authors (M.K.) acknowledges the support
by the Grants-in-Aid for Scientific Research on Innovative Areas from MEXT
(Grant No. 2404:24105008) and JSPS KAKENHI Grant 563 No. 25400240.

\end{document}